\documentclass{sig-alternate-05-2015}
\usepackage{tabu}
\usepackage{algorithm}  
\usepackage{algorithmic}  
\usepackage{booktabs}
\begin{document}

\setcopyright{acmcopyright}

\doi{10.475/123_4}

\isbn{123-4567-24-567/08/06}

\conferenceinfo{WWW 2017}{April 3--7, 2017, Perth, Australia}

\acmPrice{\$15.00}

%

\title{Question Retrieval for Community-based Question Answering via Heterogeneous Network Integration Learning}

%
%
%
%
%

\numberofauthors{4} 
%
\author{
%
%
\alignauthor
Zheqian Chen\\
       \affaddr{State Key Lab of CAD \& CG}\\
       \affaddr{Zhejiang University}\\
       \email{zheqianchen@gmail.com}
\alignauthor
Chi Zhang\\
       \affaddr{State Key Lab of CAD \& CG}\\
       \affaddr{Zhejiang University}\\
       \email{wellyzhangc@zju.edu.cn}
\and  
\alignauthor 
Zhou Zhao\\
       \affaddr{Key Lab of DCD}\\
       \affaddr{Zhejiang University}\\
       \email{zhaozhou@zju.edu.cn}
\alignauthor 
Deng Cai\\
       \affaddr{State Key Lab of CAD \& CG}\\
       \affaddr{Zhejiang University}\\
       \email{dengcai@gmail.com}      
}



\maketitle

\begin{abstract}
Community-based question answering platforms have attracted substantial users to share knowledge and learn from each other. As the rapid enlargement of CQA platforms, quantities of overlapped questions emerge, which makes users confounded to select a proper reference. It is urgent for us to take effective automated algorithms to reuse historical questions with corresponding answers. In this paper we focus on the problem with question retrieval, which aims to match historical questions that are relevant or semantically equivalent to resolve one's query directly. The challenges in this task are the lexical gaps between questions for the word ambiguity and word mismatch problem. Furthermore, limited words in queried sentences cause sparsity of word features. To alleviate these challenges, we propose a novel framework named \textbf{HNIL} which encodes not only the question contents but also the asker's social interactions to enhance the question embedding performance. More specifically, we apply random walk based learning method with recurrent neural network to match the similarities between asker's question and historical questions proposed by other users. Extensive experiments on a large-scale dataset from a real world CQA site Quora show that employing the heterogeneous social network information outperforms the other state-of-the-art solutions in this task. 
\end{abstract}

\keywords{CQA; Question retrieval; Deep learning; Social network}

\section{Introduction}
Community-based question answering (CQA) services enable users to put forward their puzzles and share knowledge with each other. Over the past years, CQA services like Yahoo! Answers, Baidu Knows, Wiki Answers, Zhihu and Quora have accumulated substantial question-answer pairs~\cite{wang2013wisdom}. However, large quantities of proposed questions are highly overlapped and redundant, which weakens users query efficiency~\cite{Lei2015Semi}. To effectively automate select the proper references from the large-scale pre-queried questions with corresponding answers, researchers have devoted into question retrieval, question answering, expert finding and natural language processing field for many years. 

In this paper we focus on the domain of question retrieval. The critical problem of question retrieval is to help users to retrieve historical questions which precisely match their questions semantically equivalent or relevant~\cite{jeon2005finding}. Users can refer to the good matches before choosing whether to raise a new question. The functionality brings users much convenience and reduces the repetition rate for CQA platforms. Hence, it is of great value for CQA services to offer relevant results efficiently and precisely. Many studies have been done on this task. However, challenges still remain due to the lexical gaps between questions caused by word ambiguity and word mismatch problem~\cite{zhou2015learning}. For example, in Quora site there exit two questions ``What are some good introductory materials on machine learning?'' and ``How can I start learning machine learning?''. From the views of our human readers, these two questions are semantically relevant and exactly express the same meaning. While for the main stream models applied in question retrieval, these two questions share few common words so they may cause the dismatch problem. Even for the same word may cause ambiguity, for instance when we mention the word `apple', we can not easily tell whether it is about the apple company or the apple fruit~\cite{De2014Towards} unless we classify through context information. Another challenge in question retrieval is the feature sparsity issue~\cite{Min2016Sparse}. As question titles usually have short length with varieties of irregular noise, it is hard to extract exactly modeling topic from using the full information of questions.  

Most of the existing works consider the question retrieval task as a supervised learning method, which utilizes both the question textual content and its belonging category to train an evaluating model~\cite{cao2012approaches},~\cite{zhou2013improving},~\cite{zou2015learning}. Researchers in question retrieval field mainly exploit the language model to learn the semantic representation of question contents. Although the existing question retrieval methods have achieved excellent performance, they do not fully tackle the word sparsity bottleneck and utilize the questions side information such as asker's background, which is critical for question understanding. Moreover, since askers have their own social network and their interests may be similar with their friends, it is reasonable to assume one scenario: Users may post questions resembled with their friends. It is a very common phenomenon among classmates and colleagues. Thus, how to leverage these available social information is of significance for the question retrieval task. 

Apart from the valuable social information, the textual contents of questions are necessary for question retrieval tasks. Recent works on question retrieval in CQA data employ different retrieval models to learn semantic representations, including the language model~\cite{zhou2014group}, the translation model~\cite{zhou2011phrase} and learning-to-rank model~\cite{zou2015learning}. Empirically these previous works consistently show the feasibility in retrieval performance. However, traditional hand-crafted features like bag-of-words have inevitable issues that can not well-embed the word suquence of questions. Inspired by the flourish of deep learning application in natural language processing~\cite{le2014distributed}, various embedding methods are proposed for learning the semantics of similar words and encode the word sequence into low-dimensional continuous embedding space. Since the question contents are always sequential data with variant length, recurrent neural network~\cite{hochreiter1997long} is an ideal choice to learn the semantic representation. 

In this paper, we put forward a novel framework named HNIL(\textbf{H}eterogeneous \textbf{N}etwork \textbf{I}ntegration \textbf{L}earning). Specifically, we exploit a random walk method to explore the valuable side information from heterogeneous social network and question category information. Besides, we model the question textual content with recurrent neural network. We then concatenate the question textual content with user embedding to represent the question and rank similarities with historical questions. In our proposed HNIL framework, the questions textual content, their related categories information, askers social information are simultaneously learned so we can utilize the rich interactions between CQA data and users data. When a new question is queried, HNIL can rank the historical proposed similar questions so that users can refer to the recommended questions along with corresponding answers without having to wait his own question to be answered. 

It is worthwhile to highlight several contributions of our work here:
\begin{itemize}
\item We introduce a novel framework named HNIL to integrate question textual content with asker social network information. We utilize a random deep walk method with recurrent neural network to learn the semantic representation of questions and users simultaneously.

\item Unlike previous studies, our proposed framework which leverages the semantic representation of questions and the rich heterogeneous social network information in question retrieval field. The framework can be extended to other information retrieval field for it is scalable for heterogeneous network learning.

\item Our proposed framework outperforms the state-of-the-art models that utilized only question textual information. The performance improved significantly in question retrieval ranking, which demonstrate the potential of our concept of integrating the rich social network side information.

\end{itemize}

The remainder of this paper is organized as follows. In Section 2, we present a brief view of current related work about question retrieval. In Section 3, we formulate the question retrieval problem and introduce our proposed heterogeneous network integration learning method. In Section 4, we describe the experimental settings and report a variety of results to varify the superiority of our model. Finally, we conclude the paper in Section5.

\section{Related Work}
The existing methods for question retrieval can be basically categorized as categories-model based approaches, translation-model based approaches, topic-modeling based approaches and neural network based approaches. 

The first approach is the most widely considered in exploring question retrieval problems. It considers the metadata of questions by taking question categories and labels into consideration. Cao et al.~\cite{cao2009use},~\cite{cao2010generalized},~\cite{cao2012approaches} embodied three language models to exploit question categories smoothing for estimating questions similarities under the same category. Zhou et al.~\cite{zhou2013towards},~\cite{zhou2014group},~\cite{zhou2015learning} proposed several methods in employing category side-information. In~\cite{zhou2013towards} they leveraged user chosen category and filter irrelevant questions under leaf categories. In~\cite{zhou2014group} they developed group non-negative matrix factorization with learning the category-specific topics for each category as well as shared topics across all categories. Zhou et al.~\cite{zhou2015learning} also employed fisher kernel to aggregate word embedding vectors from variable size into fixed-length, thus learnt a continuous word embedding model. 

The second approach, translation-model based method learns the pair relevance of question-answer data to bridging Lexical gaps between queries and questions, or questions and answers. Jeon et al.~\cite{jeon2005finding} discussed a method that refers to the similarities between answers to estimate question semantically similar probabilities. Xue et al.~\cite{xue2008retrieval} combined the question part with a query likelihood approach by incorporating word-to-word translation probabilities. Lee et al.~\cite{lee2008bridging} investigated empirical methods to eliminate non-topical or unimportant words in order to construct compact translation models for retrieval purposes. Apart from word-level translation method, Zhou et al.~\cite{zhou2011phrase} learnt a phrase-based translation model which aims to capture question contextual information rather than word-based in isolation. 

The third approach topic-modelling based approaches also arised many attentions for we can compare the latent similarity of questions without being constrainted by the queried sentences forms. Duan et al.~\cite{duan2008searching} identified the question topic and focus into a consisting data structure. Zhang et al.~\cite{zhang2014question} assumed that questions and answers share some common latent topics and through this way the model can match questions on a topic level. Chen's et al.~\cite{ji2012question} assumption is quite similar to Zhang et al~\cite{zhang2014question}.

The fourth approaches leverage neural network to model questions embeddings. As the flourish of deep learning especially in natural language processing, researchers bagan to incorporate the neural network into learning to rank frameworks. Zhou et al.~\cite{zhou2016learning} learnd the semantic representation of queries and answers by using a neural network architecture. Although the mainstream models of neural network are mainly applied in question answering not in question retrieval, the theories are the same. Qiu et al.~\cite{qiu2015convolutional} encoded questions and answers in semantic space and model their interactions in a convolutional neural tensor network architecture. The model is a general architecture with no need for lexical or syntactic analysis. Shen et al.~\cite{shen2015question} utilized a similarity matrix which contains both lexical and sequential information to effectively model the complicated matching relations between questions and answers.

In question retrieval field, very few approaches consider the heterogeneous social network to dig more information. For example, classmates or colleagues concern the same professional field so they may care about the similar questions. Zhao et al.~\cite{Zhao2015Expert} implemented a graph-regularized matrix completion algorithm by integrating the user model to improve expert finding performance. The cross-domain social information integration is also considered in Jiang et al.~\cite{Jiang2015Social}. They proposed a star-structured hybrid graph centered network in a social domain and utilized random walk method to predict user-item links in a target domain. Although these methods exploit social information contained in the social link structures, they treat the items (e.g., questions and answers) and users as simple nodes and ignore the rich content information. Theoretically, matching the similarity of queried questions via heterogeneous social network can make a better use of question related information. 

\section{Question Retrieval via Heterogeneous Network Ranking Learning}

In this section, we formulate the question retrieval problem, and propose our HNIL framework with details, finally we introduce how to use a random walk along with recurrent neural network to train our model.

\subsection{The Problem}
We consider the problem of question retrieval from the view point of integrating heterogeneous social network with question content textual information. We first denote the questions with proper semantic embeddings. Since the questions proposed are variable length of sequential data, we construct recurrent neural networks to encode the question textual content into fixed length feature vectors. Recurrent neural network have shown great superiority in dealing with variable word sequence length~\cite{le2014distributed}. Given a sequence of questions $X=\left \{ \chi _{1},\chi _{2},\chi _{3},...,\chi _{n}\right \}$, we represent the question words by word2vec embedding pre-trained by Mikolov et al.~\cite{Mikolov2013Efficient} and then obtain the latent semantic embedding with fixed length feature vectors from recurrent neural network, which denote as $Q=\left \{ q_{1},q_{2}, q_{3},...,q_{n}\right \}$. In our model, we exploit not only the question textual information, but also the side information of askers social interaction. We give our assumption of the question information in Figure 1.

\begin{figure}[t]
                        \centering
                        
                        \includegraphics[width=7cm]{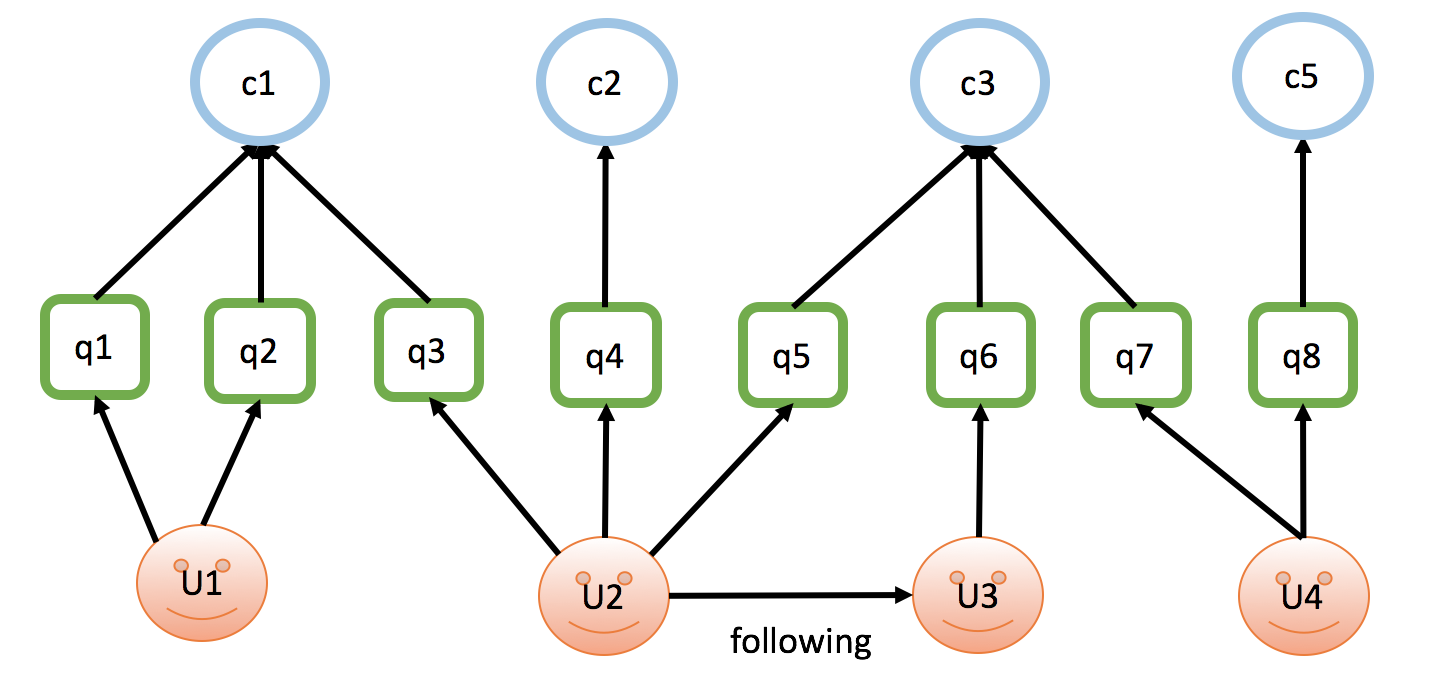}
                        \caption{Heterogeneous CQA Network, in the graph c denotes as category, q denotes as question, u denotes as user}
                    \end{figure}
Figure 1 shows our assumption of the heterogeneous social networks of askers, their proposed questions and corresponding categories. We hold the confidence that the inner differences between questions in the same categories are less than those not in the same one. What's more, friends are someone who share common interests thus they may concern extremely similar questions. The reflecting result on question retrieval is an asker may query some questions which are similar to what his friends once queried. Our method is mainly based on these two assumptions. As we can see in Figure 1, $(q2, q3)$ under the same category have a high probability that are more similar to $(q3,q4)$ for $q3$, $q4$ belong to distinct categories. Besides, we can see another three questions in category 3, $(q5, q6)$ are proposed by $(u2, u3)$ respectively, and $(u2, u3)$ are friends, while $u4$ has no relationship with $u2,u3$. So we can infer that $(q5, q6)$ are more similar than $q7$ since friends may concern the same problem.

We employ the ranking metric function $f_{q_{i}}\left ( q_{j} \right )=q_{j}^{T}q_{i}$ that quantifies the similarities between question i and question j. We denote a triplet constraint by the ordered tuple $\left( j,i,k \right )$, meaning that ``the $j$th question is more similar with question i compared with question k''. Let $T=\left \{ \left ( j,i,k \right ) \right \}$ denote the set of triplet constraints, we define that questions proposed in the same category or proposed by two friends get higher similarity score than others. Specifically, we can calculate the match score for a pair of questions by 
 \begin{equation}
             s\left ( q_{1},q_{2} \right )=q_{1}^{T}q_{2}  
 \end{equation}

 And we aim to learn the ranking metric function that for any $\left(j,i,k \right)\in T$, the inequality represents:
 \begin{equation}
      f_{q_{j}}\left ( q_{i} \right )> f_{q_{k}}\left ( q_{i} \right )\Leftrightarrow q_{i}^{T}q_{j}>q_{i}^{T}q_{k} 
 \end{equation}

 Many community question answering websites require users to offer their social accounts, we observe that the questions proposed by users have centain relevance with users attributions and their social relationships. Thus we construct a relation matrix which denotes users relationships by $M\in R^{m\times m}$. We let the entry $s_{ij}=1$ if the i-th user  and the j-th user are friends, otherwise, $s_{ij}=0$. We then integrate the question content textual information along with user social network information to rank the similarities between questions. In this way we get the heterogenous CQA networks to tackle the sparsity bottleneck of question data. Formally speaking, for a network $G=\left(V,E \right)$, where $V$ is the set of nodes denoting questions and users. $E\subseteq (V\times V)$ is the set of edges in $G$ consists of question relation and user social relation, we want to learn the latent semantic embeddings for each node in $V$. Figure 2 guides us ideas on how to concate questions and users synchronously to learn representations of contexts information.

 \begin{figure}[t]
                        \centering
                        
                        \includegraphics[width=7cm]{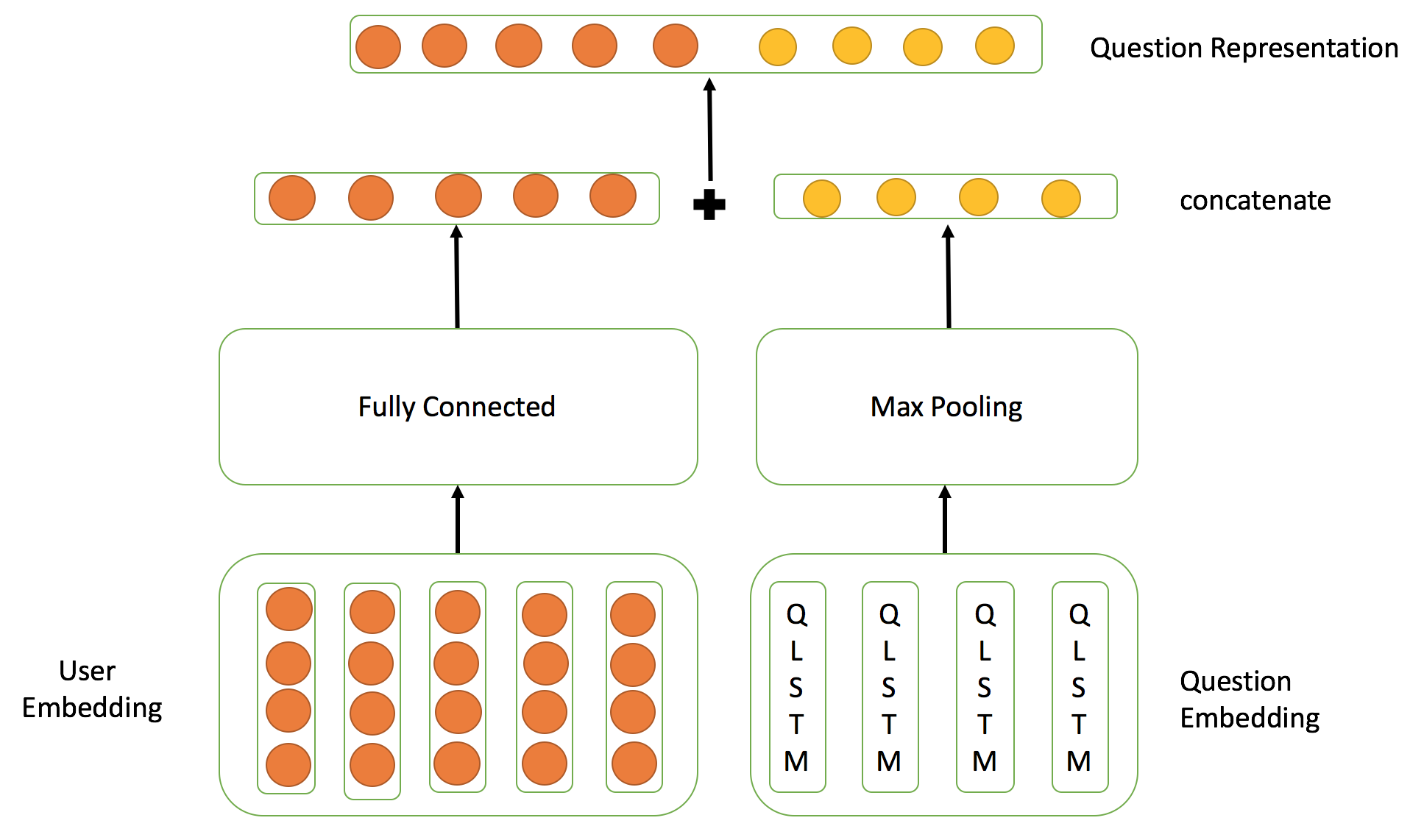}
                        \caption{User modeling from a learned user embedding matrix and question modeling with LSTM. We concatenate these two embedding results into a unified representation of question.}
                    \end{figure}

 As Figure 2 illustrated, the input of context includes users embedding and questions embedding. We learn user representation $f(u)$ from a learned user embedding matrix and apply LSTM to encode questions. Then we concatenate two vector embeddings into a unified vector. We regard the concatenated vector as the updated question representation accompanied by asker information. To this end, the vector of each context word is made up of two parts: a user embedding $e_w=L_wi_w$ and question embedding, where $L_w\in M^{d_w\times V_w}$ is the lookup tabels of user representations, $d_w$ and $V_w$ are the dimensions of word vector and user numbers.

 Formally, we elaborate the question retrieval problem by using the above notations as follows. For a set of existed questions, given a new queried question $X$ along with its asker information, we learn the latent representation of this new queried question through heterogeneous CQA networks so that we can retrieve the top-n most similar questions in our database. For fomulation, we learn the heterogeneous CQA networks of graph $G$, and then embed the askers $u$ and questions $q$ to rank the the similarity $f_{q_{j}}(q_{i})$. 

\subsection{Heterogeneous Network Ranking Learning with Recurrent Neural Networks}
In this section, we introduce the Heterogeneous network ranking learning framework with recurrent neural network for question retrieval.

In CQA service, there exits a signicant distinction that users have abundant interaction in social networks. For example, users may concern some certain fields and some field experts, or they share knowledge with friends and may concern some questions that were asked by friends. Thus, it naturally forms the heterogenous CQA network between questions and users. As we have shown in Figure 1, there exits two network：one network is the pure attributes of questions and their corsesponding categories, another network is social relationship network whose nodes are users. 

In this paper, we consider how to exploit the rich interaction information from heterogenous CQA network. Inspired by DeepWalk method proposed by Perozzi et al.~\cite{perozzi2014deepwalk}, we use deep random walk to combine the blended nodes into mutiple paths, and then we consider the sampled paths as the context windows for the vertex embedding in networks. we treat the paths as sentences and each node $v_{i}$ in paths regarded as words. In traditional language modelling, given a sequence of words, we aim to estimate the likelihood of these specific words in the whole oracle. Recently probabilistic neural networks fucus on extending the traditional language model to generalize its original target. More formally in our framework, we explore the question graph through a series of short random walks to build language modeling generalization. These paths can be regarded as short sentences and phrases in a special language. So given all the previous vertices visted, the direct intention is to evaluate the likelihood of observing vertex $v_i$ in the walked path. Our goal is to learn the posterior distribution which we can predict the context node $v_{i}$ within a window $W=\left \{v_{i-w},v_{i-w+1},...,v_{i+w-1},v_{i+w} \right \}$ by SkipGram. Following the neural language models by Mikolov et al.~\cite{Mikolov2013Efficient}, we formulate the vetex as the optimization problem as follows:
\begin{equation}
\min_{\phi} -logPr\left ( \left \{ v_{i-w},...,v_{i-1},v_{i+1},...,v_{i+w} \right \} \mid \phi \left( v_{i}\right)\right )
\end{equation}
where $\phi$ represents the latent embedding of nodes V and $\phi \left(v_{i}\right)$ represents the latent embedding of $v_{i}$. Compared to traditional latent representation learning of probability distribution of node co-occurrences, the traditional likelihood $Pr(v_i|(\phi (v_1),(\phi (v_2),...,(\phi (v_{i-1})))$ is unfeasible when the walk length grows and  the computing cost will be huge. As for the optimization method proposed by Mikolov et al.~\cite{Mikolov2013Efficient}, it uses one word to predict the context instead of using context to predict a missing word. The words appearing in both left and right hand of given words compose the context so that remit the ordering constraint on language modeling. We find the optimization function is particularly desirable for social representation learning. The advantages of this optimization method are the speed up for training time by building small models as one vertex is given at a time, besides the order independence assumption better captures the sense of nearness, which is criticle for short text matching task.

The original deepwalk proposed by Perozzi et al.~\cite{perozzi2014deepwalk} is an unsupervised learning framework which only learns nodes embedding from structured graph. Yet in our proposed HNIL framework, the question side information we exploit are question contents as well as their corresponding categories and askers embedding. It is critical to leverage supervised learning method to integrate these available information. Thus we need to adapt the original deepwalk framework to fulfill our model. Naturally, since questions content involve word sequence learning, we integrate the deepwalk method with deep recurrent neural network learning into a unified CQA network learning framework to boost question retrieval performance. 

We then describe our approaches for learning question embeddings and user representations. To begin with we learn the question representation through recurrent nerual network. We choose long-short term memory ~\cite{hochreiter1997long} instead of traditional recurrent neural network to learn question embeddings. For traditional recurrent neural network lack the ability of controling long term dependencies as mentioned by Sutskever et al.~\cite{sutskever2014sequence}. We learn the embeddings of questions by following equations:
\begin{align*}
 i_{t}&=\delta (W_{i}x_{t}+G_{i}h_{t-1}+b_{i}) \\
\hat{C}_{t}&=tanh(X_cx_t+G_fh_{t-1}+b_f) \\
f_t&=\delta (W_fx_t+G_fh_{h-1}+b_f) \\
C_t&=i_t\cdot \hat{C_t}+f_t\cdot C_t\\
o_t&=\delta(W_ox_t+G_oh_{t-1}+V_oC_t+b_o)\\ 
h_t&=o_t \cdot tanh(C_t)
\end{align*}

where $\sigma$ represents the sigmoid activation function; $W_s$, $U_s$ and $V_o$ are weight matrices; and $b_s$ are bias vectors. There are three different gates (input, output, forget gates) for controlling memory cells and their visibility. The input gate can allow incoming signal to update the state of the memory cell or block it and the output gate can allow the state of the memory cell to have an effect on other neurons or prevent it. Moreover, the forget gate decides what information is going to be thrown away from the cell state. We take the output of the last LSTM cell, $h_k$, as the semantic embedding of the input sequence $\left \{x_1,x_2,...,x_k \right \}$.

Since questions are often queried as paragraphs with several sentences, we split the question paragraphs into single sentence so we can learn the semantic embedding by LSTM as described above, we then merge the output embeddings of LSTM by adding an additional max-pooling layer.

As for user embedding, we first construct a user social matrix $M\in R^{m\times m}$, we let each entry $s_{ij}=1$ if the i-th user  and the j-th user are friends, otherwise, $s_{ij}=0$ to denote users social relationships. And then we learn user embedding matrix and get a fully connected layer.

We illustrate our framework in Figure 3. 

\begin{figure*}[t]
                        \centering
                        
                        \includegraphics[width=15cm]{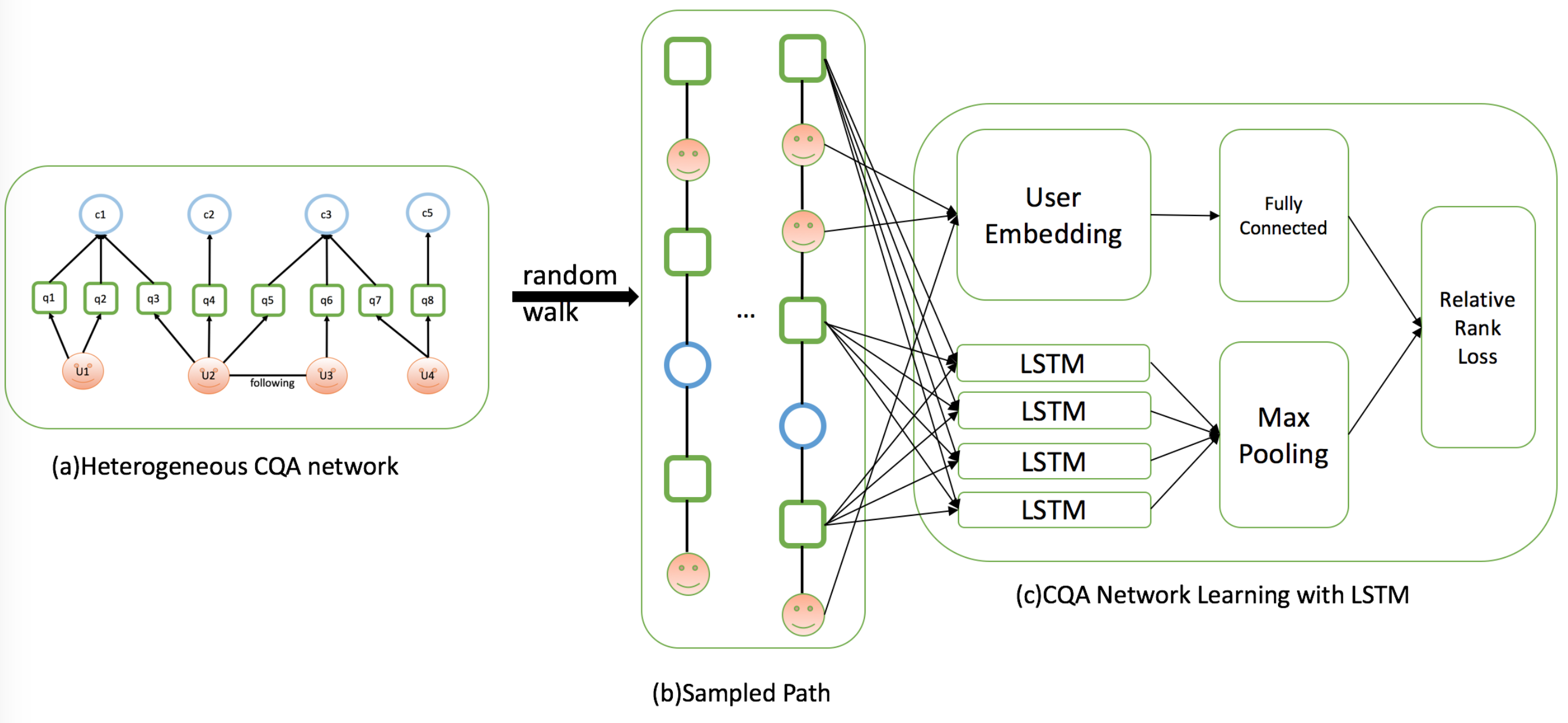}
                        \caption{The Overview of Heterogeneous Network Ranking Learning. (a) The heterogeneous CQA network is constructed by integrating questions content as well as corresponding categories and askers social network relation. (b) A deep random walker is walking on the heterogeneous CQA networks to sample the data paths. (c) The questions, askers and categories are encoded into fixed feature vectors by specific models, these features are used to calculate the matching score between questions as output in testing process or obtain loss to update the parameters in training process. }
                    \end{figure*}

As we can see in Figure 3, we construct the heterogeneous CQA network and apply a deep random walk to sample the node path, and then we learn the nodes representations with LSTM respectively. Taking notice of that the framework involves three different nodes and to utilize the rating information, for each node $v_i$ in context window $W$, we design a specific loss function to simultaneously rank the relative similarities between different questions. The loss function is designed as follows:
\begin{equation}
l(v_i)=\left\{\begin{aligned}
&\sum _{q+,q-\in W}max(0,m+f_{q-}(q_i)-f_{q+}(q_i)), q_i\in Q   \\ 
&\sum_{u \in W}\left \| u-v_i \right \|,q_i\in U \\ 
\end{aligned}\right.
\end{equation}

Where the superscript $q+$ denotes the higher-score of similar question and $q-$ denotes the lower-score of similar question. In our task, we let questions under the same category denoted as $q+$ while questions under diverse categories denoted as $q-$. We denote the hyper-parameter m $(0<m<1)$ controls the margin in hinge loss function and $C,U$ are the sets of categories and askers, respectively. We integrate the heterogeneous CQA networks by using the framework described above to learn the question retrieval analysis.  

\subsection{Heterogeneous Network Ranking Learning}  
In this section, we describe the detail of learning our proposed framework and suggest the heuristic approaches in adopting heterogeneous social information to enhance the model performance. 

We incorporate the question textual contents and askers' relative relationships into a unified CQA network embedding framework for question retrieval. Integrating users' social information can help tackle the question words sparsity bottleneck and adopt more side information for understanding the question intent, which is critical for question similarity retrieval. The trained model can rank the question similarities for given a question directly. 

We then summarized the main training process algorithm in Algorithm 1.

\begin{algorithm}
\caption{HNIL for question retrieval in CQA}
\begin{algorithmic}[1]
\REQUIRE ~~\\ 
The set of heterogeneous network $G(V,E)$, Windows size $w$; Max walk length $T$, Number of iterations $m$, Walks per vertex $n$, embedding size $d$\\    
\STATE Pre-train nodes embedding matrix by Deepwalk
\FOR{$t=1$ to $T$} 
   \FOR{$j=1$ to $n$} 
        \STATE$O=Shuffle(V)$
        \FOR{$V \in O$} 
        \STATE $p=Deepwalk(G,v,t)$
        \STATE for q node: q=lstm(q).concatenate v(u)
        \STATE calculate the loss for each node in $p$
   \ENDFOR
   \STATE Summate the total training loss
   \STATE Update parameters by SGD
   \ENDFOR
\ENDFOR
\end{algorithmic}
\end{algorithm}

The framework is illustrated via Figure 3. Before training across the heterogeneous CQA network, we sample the node paths through a deep random walker. We assign the category node to be the supervised signal and learn the representation of question nodes and user nodes. The loss of each node is accumulated to get the final training loss. We manage to minimize the total loss by using stochastic gradient descent(SGD) with diagonal variant of AdaGrad~\cite{qiu2015convolutional}. 

Denote all the parameters in our proposed framework as $\Theta$, our learning process is given by:
\begin{equation}
\min_{\Theta} L(\Theta)=\sum _p\sum _{v_{i}}l(v_i)+\lambda \left \| \Theta \right \|_{2}^2 
\end{equation}
where $\lambda>0$ is a hyper-parameter that trades off between the training loss and regularization. 

By using SGD optimization, at time step t, the parameter $\Theta$ is updated as follows:
\begin{equation}
\Theta_t=\Theta_{t-1} - \frac{\rho }{\sqrt{\sum_{i=1}^{t}g_i^2}}g_t
\end{equation}

where $\rho$ is the initial learning rate and $g_t$ is the subgradient at time t.

\section{Experiments}
In this section, we present experiments to evaluate the performance of the proposed method for question retrieval. The experiments are based on the Quora service and the twitter user social network. 

\subsection{Data Preparation}
We collect the data sets from the community question answering service Quora and askers' following relationships in twitter social network. The dataset includes 50451 questions, 4415 users and their following relationships. Each question in this collected corpus consists of three fields: question content information and its corresponding tag, as well as the asker's twitter id and his following friends' id. The dataset is splited into training set and testing set without overlapping. The size of test set is fixed as $20\%$ and the size of training set varies from $20\%$ to $80\%$. 

\subsection{Evaluation Criteria}
In order to evaluate the performance of different models, we employ Mean Average Precision (MAP), Precision at K (P@k) and Mean Reciprocal Rank (MRR) as evaluation measures. These measure criterions are widely used in the evaluation for question retrieval performance in CQA. MAP reports the model relevance ranking ability for a ranked sequence of documents. Precision@K is a useful metric corresponds to the number of relevant results on top k search results. MRR evaluates a sequence of possible responses to a sample of queries, ordered by probability of correctness. 

We now introduce the evaluation criterias in details below.
\begin{itemize}
\item \textbf {MAP (Mean Average Precision)} for a set of queried questions Q is the mean of the average precision scores for each query for a method M:
\begin{align}
& Map=\frac{\sum _{q\in Q}AvgP\left ( q \right )}{\left | Q \right |}  \label{eq:rel1} \\
& AvgP\left ( q \right )=\frac{1}{N_{M_{q}}}\sum_{\left | M_{q} \right |}^{j=1}\frac{N_{M_{q,j}}}{j}\mathbf{1}\left ( M_{q,j} \right )  \label{eq:rel2}
\end{align}

where $1(S)$ is a binary function indicates the relevance denotes as 1 if $M_{q,j}$ is relevant and 0 as irrelevant. $N_{M_{q,j}}$ is the number of relevant questions among the top $j$ ranked list, and $N_{M_{q}}$ denotes the total number of relevant questions of queried question $q$, and $M_{q,j}$ is the $j$th question in the sequence list for queried question.

\item \textbf {P@N (Precision @N)} for a set of queried questions Q, P@N measures the ratio of the top N retrieved questions that are relevant to the queried questions for a method M:
$$
p@N=\frac{1}{\left | Q \right |}\sum_{q\in Q}\frac{N_{M_{q,N}}}{N}
$$

where $N_{M_{q,N}}$ denotes the number of relevant questions given top $N$ ranked list returned for queried question $q$.

\item \textbf {MRR (Mean Reciprocal Rank)} MRR is broadly used in multi-results problems. In some specific problems, if only return the top 1 result, the precision rate or recall rate will be poor, so we usually return multi-results if the technique is not ripe. The core idea of MRR is simple: the merits of models related to the location of the first correct result, the earlier, the better. The mean reciprocal rank is the average of the reciprocal ranks of results for a set of queries Q:
$$
MRR=\frac{1}{\left | Q \right |}\sum_{i=1}^{\left | Q \right |}\frac{1}{rank_{i}}
$$

Where $\left | Q \right |$ denotes total number of query sets and $rank_i$ denotes the sequence location of the correct result.

\end{itemize}

\subsection{Performance Comparisons}
To validate the performance of our approach, we compare our proposed method against with other six state-of-the-art methods for the problem of question retrieval in CQA site as follows:

\begin{itemize}
\item \textbf{VSM} Vector Space Model is an algebraic model for representing text documents as vectors of identifiers. In our experiment, we represent questions by VSM feature vectors and then calculate the relevant score to rank the similarites between questions.
\item \textbf{BM25} Okapi BM25 proposed by Stephen E et al.~\cite{Robertson1994OkapiAT} (BM stands for Best Matching) is a ranking function used to rank matching documents according to their relevance to a given search query. In our experiment, we exploit the original BM25 to rank a set of questions similarities based on the query terms appearing in each question.
\item \textbf{Doc2Vec} Doc2Vec (Le and Mikolov et al.)~\cite{le2014distributed} modifies the word2vec algorithm to unsupervised learning of continuous representations for larger blocks of documents. In our experiments we encode the questions as documents in a low-dimensional continuous feature space and then conduct the similarities ranking in this learned feature space.
\item \textbf{DRLM} DRLM (Kai Zhang et al.)~\cite{Zhang2016Learning} simultaneously learn vectors of words and vectors of question categories by optimizing an objective function. In experiments, we incorporate learnt representations into traditional language models. 
\item \textbf{RCNNs} RCNNs proposed by Tao Lei et al.~\cite{Lei2015Semi} apply a recurrent and convolutional model(gated convolution) to effectively map questions to their semantic representations. In our experiments, we pre-train question content with category section within an encoder-decoder framework and then use RCNNs to rank the task.
\item \textbf{DeepWalk} DeepWalk ~\cite{perozzi2014deepwalk} proposed by Perozzi et al. learns the representations of heterogeneous data depending on sampling the graph structure information of social networks.

Among these six baselines, the VSM and BM25 methods are the traditional algorithms used in information retrieval and learn the question model only based on bag-of-words contents. The Doc2Vec method is the modified version of word2vec to learn the continuous representations in low-dimensional space and can be used in larger block of texts such as sentences, paragraphs and even the entire document. These three methods focus on the representation learning of question contents to construct a semantic feature space. While the DRLM and RCNNs methods learn the question model based both on the question contents and corresponding categories. And the last DeepWalk method learn the question model based on question contents and categories can be regarded as a simplified HNIL without the extensions of user social network learning. Unlike the previous studies, our method HNIL learns the question model from the Heterogeneous CQA network along with social network. The input words of our methods are initialized by pre-trained word embeddings~\cite{Mikolov2013Efficient} and the weights of LSTMs are randomly by a Gaussian distribution with zero mean. We then employ the random-walk based learning with LSTM networks for training our proposed HNIL model.       
\end{itemize}

\subsection{Experimental Results and Analysis}
To evaluate the performance of our proposed framework, we conduct several experiments on four metrics described above.

Table 1, 2, 3 and 4 show the evaluation results on MAP, Precision@1, Precision@5 and MRR, respectively. We conduct the experiments by separating the whole data into different ratio from 20\%, 40\%, 60\% to 80\%. We choose the parameters which achieve the best performance to implement the testing evaluation. We then report several interesting analysis that we observed on the evaluation results .  
\begin{table}[htbp]
 \caption{\label{tab:test}Experimental results on MAP with different proportions of data for training.(best scores are boldfaced)}
 \centering
 \begin{tabular}{lclccccl}
  \toprule

  Training& 20\% & 40\% & 60\% & 80\% \\
  \midrule
 VSM      &0.1771 &0.1893 &0.2022 &0.2275 \\       
 BM25     &0.1404 &0.1521 &0.1393 &0.1583 \\      
 Doc2Vec  &0.2153 &0.2247 &0.2336 &0.2568 \\      
 DRLM     &0.2322 &0.2508 &0.2819 &0.3117 \\       
 RCNNs    &0.2111 &0.2312 &0.2494 &0.2689 \\       
 DeepWalk &0.1825 &0.2029 &0.2150 &0.2281 \\
 HNIL     &\textbf{0.3321} &\textbf{0.3723} &\textbf{0.3764} &\textbf{0.4067} \\    

  \bottomrule

 \end{tabular}
\end{table}

\begin{table}[htbp]
 \caption{\label{tab:test}Experimental results on Precision@1 with different proportions of data for training.(best scores are boldfaced)}
 \centering
 \begin{tabular}{lclccccl}
  \toprule

  Training& 20\% & 40\% & 60\% & 80\% \\
  \midrule
 VSM      &0.1733 &0.1821 &0.2027 &0.2262 \\   
 BM25     &0.1600 &0.1820 &0.2215 &0.2503 \\ 
 Doc2Vec  &0.1898 &0.2231 &0.2343 &0.2748 \\  
 DRLM     &0.2131 &0.2449 &0.2780 &0.3207 \\
 RCNNs    &0.1909 &0.2353 &0.2522 &0.2801 \\  
 DeepWalk &0.1728 &0.2197 &0.2635 &0.2955 \\
 HNIL     &\textbf{0.3341} &\textbf{0.3551} &\textbf{0.3867} &\textbf{0.3993} \\
 \bottomrule

 \end{tabular}
\end{table}

\begin{table}[htbp]
 \caption{\label{tab:test}Experimental results on Precision@5 with different proportions of data for training.(best scores are boldfaced)}
 \centering
 \begin{tabular}{lclccccl}
  \toprule

  Training& 20\% & 40\% & 60\% & 80\% \\
  \midrule
 VSM      &0.1589 &0.2073 &0.2353 &0.2709 \\  
 BM25     &0.1240 &0.1552 &0.1974 &0.2231 \\    
 Doc2Vec  &0.1919 &0.2117 &0.2908 &0.3275 \\       
 DRLM     &0.2447 &0.2692 &0.3046 &0.3681 \\
 RCNNs    &0.1936 &0.2499 &0.3051 &0.3269 \\
 DeepWalk &0.1946 &0.2363 &0.2727 &0.3021\\
 HNIL     &\textbf{0.2962} &\textbf{0.3093} &\textbf{0.3587} &\textbf{0.4101}  \\ 

  \bottomrule

 \end{tabular}
\end{table}

\begin{table}[htbp]
 \caption{\label{tab:test}Experimental results on MRR with different proportions of data for training.(best scores are boldfaced)}
 \centering
 \begin{tabular}{lclccccl}
  \toprule

  Training& 20\% & 40\% & 60\% & 80\% \\
  \midrule
 VSM      &0.1929 &0.2071 &0.2504 &0.2992 \\           
 BM25     &0.2540 &0.2932 &0.3327 &0.3561 \\
 Doc2Vec  &0.2731 &0.3036 &0.3791 &0.3903 \\              
 DRLM     &0.3273 &0.3686 &0.4157 &0.4696 \\           
 RCNNs    &0.2851 &0.3109 &0.3818 &0.4054 \\              
 DeepWalk &0.3007 &0.3546 &0.4039 &0.4329 \\
 HNIL     &\textbf{0.4038} &\textbf{0.4391} &\textbf{0.4790} &\textbf{0.5322} \\     

  \bottomrule

 \end{tabular}
\end{table}

\begin{itemize}
\item The supervised methods HNIL, DRML, RCNNs, DeepWalk outperforms the unsupervised VSM, BM25 and doc2vec methods, which suggests that in question retrieval problem it is critical to use supervised information such as question category to enhance the inner similarities of questions under the same category. 
\item Since DeepWalk performs worse than DRLM in most cases, we can conclude that the content analysis plays a more import role than the simple utilization of graphical node information in question retrieval tasks. 
\item The experimental results in all cases show that our proposed HNIL achieves the best performance. The fact suggests that integrating question context under category supervised signal with user social network information can further improve the performance of question retrieval. So, in the future study we can consider more question side information to fully dig question attributions to enhance question retrieval ability. 

\end{itemize}

There are three essential parameters in our proposed framework: the length of sampled paths, the size of network embeddings and the number of walks. From our empirical settings, we vary number of walks from 5 to 20 and the dimension of embeddings from 100 to 1000, the performance trends turn stable when the number enlarged over 10 and the dimension over 200. We then analyze the effect of sampled paths length on HNIL by varying length using $60\%$ of the data, and when we use $70\%, 80\%$ of the data the trend is consistent. From Figure 4 we notice when the length of sampled paths is set to 6 we get the best performance in MAP, P@1, P@5, MRR respectively. 

\begin{figure}[t]
                        \centering
                        
                        \includegraphics[width=7cm]{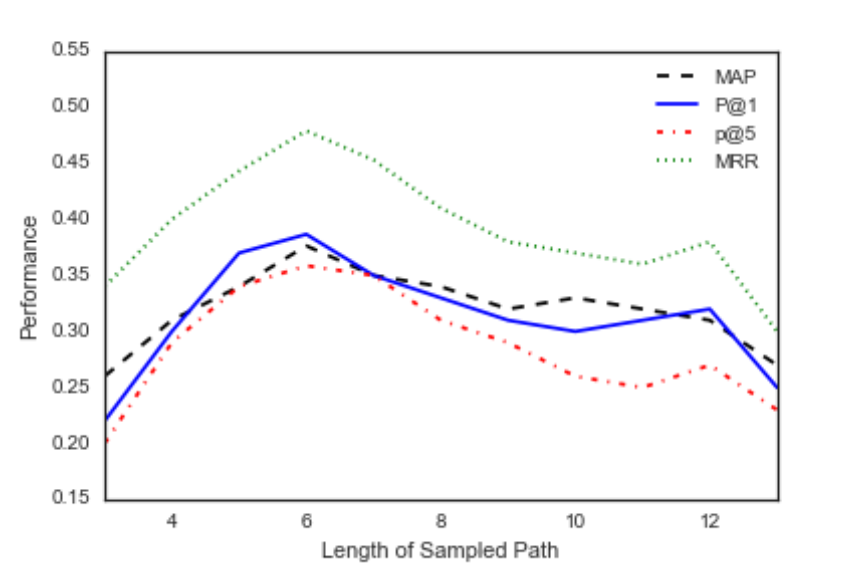}
                        \caption{Effect of Sampled Path Length on MAP, P@1, P@5, MRR using $60\%$ of data.}
                    \end{figure}

\section{Conclusion}
Question retrieval is an essential component in Community Question Answering(CQA) services. In this paper we discuss a new framework which is capable of exploiting category information associated with askers social attributions to enhance the question content embedding ability. We develop a random-walk based method with recurrent neural network for ranking question similarities in heterogeneous CQA networks. Our approach can be applied easily to existing question retrieval models and extended into other information retrieval field. Experiments conducted on a large CQA data set from Quora and Twitter demonstrate the effectiveness of the proposed technique.

This work opens to several interesting directions for future work. First, it is of relevance to apply the proposed technique to other question retrieval approaches and even other information retrieval fields. Second, associating with heterogeneous social network may be exploited to further improve the performance of our proposed model. Finally, it is of interest to explore other question side information and combining global relevance scores and local relevance scores for us to enhance the performance.

\bibliographystyle{abbrv}
\bibliography{sigproc}  

\begin{thebibliography}{10}

\bibitem{cao2010generalized}
X.~Cao, G.~Cong, B.~Cui, and C.~S. Jensen.
\newblock A generalized framework of exploring category information for
  question retrieval in community question answer archives.
\newblock In {\em Proceedings of the 19th international conference on World
  wide web}, pages 201--210. ACM, 2010.

\bibitem{cao2012approaches}
X.~Cao, G.~Cong, B.~Cui, C.~S. Jensen, and Q.~Yuan.
\newblock Approaches to exploring category information for question retrieval
  in community question-answer archives.
\newblock {\em ACM Transactions on Information Systems (TOIS)}, 30(2):7, 2012.

\bibitem{cao2009use}
X.~Cao, G.~Cong, B.~Cui, C.~S. Jensen, and C.~Zhang.
\newblock The use of categorization information in language models for question
  retrieval.
\newblock In {\em Proceedings of the 18th ACM conference on Information and
  knowledge management}, pages 265--274. ACM, 2009.

\bibitem{De2014Towards}
T.~De~Nies, C.~Beecks, W.~De~Neve, T.~Seidl, E.~Mannens, and V.~D.~W. Rik.
\newblock Towards named-entity-based similarity measures: Challenges and
  opportunities.
\newblock In {\em The Workshop on Exploiting Semantic Annotations in
  Information Retrieval}, pages 9--11, 2014.

\bibitem{duan2008searching}
H.~Duan, Y.~Cao, C.-Y. Li, n, and Y.~Yu.
\newblock Searching questions by identifying question topic and question focus.
\newblock In {\em ACL}, pages 156--164, 2008.

\bibitem{hochreiter1997long}
S.~Hochreiter and J.~Schmidhuber.
\newblock Long short-term memory.
\newblock {\em Neural computation}, 9(8):1735--1780, 1997.

\bibitem{jeon2005finding}
J.~Jeon, W.~B. Croft, and J.~H. Lee.
\newblock Finding similar questions in large question and answer archives.
\newblock In {\em Proceedings of the 14th ACM international conference on
  Information and knowledge management}, pages 84--90. ACM, 2005.

\bibitem{ji2012question}
Z.~Ji, F.~Xu, B.~Wang, and B.~He.
\newblock Question-answer topic model for question retrieval in community
  question answering.
\newblock In {\em Proceedings of the 21st ACM international conference on
  Information and knowledge management}, pages 2471--2474. ACM, 2012.

\bibitem{Jiang2015Social}
M.~Jiang, P.~Cui, X.~Chen, and F.~Wang.
\newblock Social recommendation with cross-domain transferable knowledge.
\newblock {\em IEEE Transactions on Knowledge \& Data Engineering},
  27(11):1--1, 2015.

\bibitem{le2014distributed}
Q.~V. Le and T.~Mikolov.
\newblock Distributed representations of sentences and documents.
\newblock {\em arXiv preprint arXiv:1405.4053}, 2014.

\bibitem{lee2008bridging}
J.-T. Lee, S.-B. Kim, Y.-I. Song, and H.-C. Rim.
\newblock Bridging lexical gaps between queries and questions on large online
  q\&a collections with compact translation models.
\newblock In {\em Proceedings of the Conference on Empirical Methods in Natural
  Language Processing}, pages 410--418. Association for Computational
  Linguistics, 2008.

\bibitem{Lei2015Semi}
T.~Lei, H.~Joshi, R.~Barzilay, T.~Jaakkola, K.~Tymoshenko, A.~Moschitti, and
  L.~Marquez.
\newblock Semi-supervised question retrieval with gated convolutions.
\newblock 2015.

\bibitem{Mikolov2013Efficient}
T.~Mikolov, K.~Chen, G.~Corrado, and J.~Dean.
\newblock Efficient estimation of word representations in vector space.
\newblock {\em Computer Science}, 2013.

\bibitem{Min2016Sparse}
M.~Peng, Q.~Xie, J.~Huang, J.~Zhu, S.~Ouyang, J.~Huang, and G.~Tian.
\newblock Sparse topical coding with sparse groups.
\newblock 2016.

\bibitem{perozzi2014deepwalk}
B.~Perozzi, R.~Al-Rfou, and S.~Skiena.
\newblock Deepwalk: Online learning of social representations.
\newblock In {\em Proceedings of the 20th ACM SIGKDD international conference
  on Knowledge discovery and data mining}, pages 701--710. ACM, 2014.

\bibitem{qiu2015convolutional}
X.~Qiu and X.~Huang.
\newblock Convolutional neural tensor network architecture for community-based
  question answering.
\newblock In {\em Proceedings of the 24th International Joint Conference on
  Artificial Intelligence (IJCAI)}, pages 1305--1311, 2015.

\bibitem{Robertson1994OkapiAT}
S.~E. Robertson, S.~Walker, S.~Jones, M.~Hancock-Beaulieu, and M.~Gatford.
\newblock Okapi at trec-3.
\newblock In {\em TREC}, 1994.

\bibitem{shen2015question}
Y.~Shen, W.~Rong, Z.~Sun, Y.~Ouyang, and Z.~Xiong.
\newblock Question/answer matching for cqa system via combining lexical and
  sequential information.
\newblock In {\em Twenty-Ninth AAAI Conference on Artificial Intelligence},
  2015.

\bibitem{sutskever2014sequence}
I.~Sutskever, O.~Vinyals, and Q.~V. Le.
\newblock Sequence to sequence learning with neural networks.
\newblock In {\em Advances in neural information processing systems}, pages
  3104--3112, 2014.

\bibitem{wang2013wisdom}
G.~Wang, K.~Gill, M.~Mohanlal, H.~Zheng, and B.~Y. Zhao.
\newblock Wisdom in the social crowd: an analysis of quora.
\newblock In {\em Proceedings of the 22nd international conference on World
  Wide Web}, pages 1341--1352. International World Wide Web Conferences
  Steering Committee, 2013.

\bibitem{xue2008retrieval}
X.~Xue, J.~Jeon, and W.~B. Croft.
\newblock Retrieval models for question and answer archives.
\newblock In {\em Proceedings of the 31st annual international ACM SIGIR
  conference}, pages 475--482. ACM, 2008.

\bibitem{Zhang2016Learning}
K.~Zhang, W.~Wu, F.~Wang, M.~Zhou, and Z.~Li.
\newblock Learning distributed representations of data in community question
  answering for question retrieval.
\newblock In {\em ACM International Conference on Web Search and Data Mining},
  2016.

\bibitem{zhang2014question}
K.~Zhang, W.~Wu, H.~Wu, Z.~Li, and M.~Zhou.
\newblock Question retrieval with high quality answers in community question
  answering.
\newblock In {\em Proceedings of the 23rd ACM International Conference on
  CIKM}, pages 371--380. ACM, 2014.

\bibitem{Zhao2015Expert}
Z.~Zhao, L.~Zhang, X.~He, and W.~Ng.
\newblock Expert finding for question answering via graph regularized matrix
  completion.
\newblock {\em IEEE Transactions on Knowledge \& Data Engineering},
  27(4):993--1004, 2015.

\bibitem{zhou2011phrase}
G.~Zhou, L.~Cai, J.~Zhao, and K.~Liu.
\newblock Phrase-based translation model for question retrieval in community
  question answer archives.
\newblock In {\em Proceedings of the 49th Annual Meeting of the Association for
  Computational Linguistics}, pages 653--662. Association for Computational
  Linguistics, 2011.

\bibitem{zhou2013towards}
G.~Zhou, Y.~Chen, D.~Zeng, and J.~Zhao.
\newblock Towards faster and better retrieval models for question search.
\newblock In {\em Proceedings of the 22nd ACM international conference on
  CIKM}, pages 2139--2148. ACM, 2013.

\bibitem{zhou2014group}
G.~Zhou, Y.~Chen, D.~Zeng, and J.~Zhao.
\newblock Group non-negative matrix factorization with natural categories for
  question retrieval in community question answer archives.
\newblock In {\em COLING}, pages 89--98, 2014.

\bibitem{zhou2015learning}
G.~Zhou, T.~He, J.~Zhao, and P.~Hu.
\newblock Learning continuous word embedding with metadata for question
  retrieval in community question answering.
\newblock In {\em Proceedings of ACL}, pages 250--259, 2015.

\bibitem{zhou2013improving}
G.~Zhou, Y.~Liu, F.~Liu, D.~Zeng, and J.~Zhao.
\newblock Improving question retrieval in community question answering using
  world knowledge.
\newblock In {\em IJCAI}, 2013.

\bibitem{zhou2016learning}
G.~Zhou, Y.~Zhou, T.~He, and W.~Wu.
\newblock Learning semantic representation with neural networks for community
  question answering retrieval.
\newblock {\em Knowledge-Based Systems}, 93:75--83, 2016.

\bibitem{zou2015learning}
Y.~Zou, T.~Ye, Y.~Lu, J.~Mylopoulos, and L.~Zhang.
\newblock Learning to rank for question-oriented software text retrieval (t).
\newblock In {\em Automated Software Engineering (ASE), 2015 30th IEEE/ACM
  International Conference on}, pages 1--11. IEEE, 2015.

\end{thebibliography}

\end{document}